\input phyzzx

\def\dplus{=\hskip-5pt \raise 0.7pt\hbox{${}_\vert$} ^{\phantom 7}}
\def\dplusup{=\hskip-5.1pt \raise 5.4pt\hbox{${}_\vert$} ^{\phantom 7}}
\def\dplus{=\hskip-4.8pt \raise 0.7pt\hbox{${}_\vert$} ^{\phantom 7}}

\def\pmb#1{\setbox0=\hbox{#1} \kern-.025em\copy0\kern-\wd0
\kern0.05em\copy0\kern-\wd0 \kern-.025em\raise.0433em\box0}

\def\cN{{\cal N}}

\def\ubx{{\underline{x}}}
\def\uby{{\underline{y}}}

\font\mybb=msbm10 at 11pt

\def\bb#1{\hbox{\mybb#1}}

\def\bZ {\bb{Z}}
\def\bR {\bb{R}}
\def\rre {{\rm Re}}
\def\iim {{\rm Im}}

\def\bC {\bb{C}}

\def\e  {\epsilon}

\hfuzz 1cm

\REF\cg{ A.. Comtet and G.W. Gibbons, {\it  Bogomolny Bounds for 
Cosmic Strings}  Nucl.Phys. {\bf B299}:719, 1988.} 
\REF\vy{ B. R. Greene, A. Shapere, C. Vafa, S-T
Yau, {\it Stringy Cosmic Strings and Noncompact Calabi-Yau Manifolds},
 Nucl.Phys. {\bf B337}:1, 1990. }
  \REF\gg{ G. W. Gibbons, M. B. Green, M. J. Perry 
 {\it Instantons and Seven-Branes in Type IIB Superstring Theory},
  Phys.Lett. {\bf B370}:37-44,1996; hep-th/9511080.}
  \REF\rom{L.J. Romans, {\it Massive N=2A Supergravity in Ten-Dimensions}
  Phys.Lett. {\bf B169}:374,1986.}
   \REF\wipo{J. Polchinski, E. Witten 
{\it  Evidence for Heterotic - Type I String Duality},
 Nucl.Phys. {\bf B460}:525-540,1996;
 hep-th/9510169.}
  \REF\ptb{E. Bergshoeff, M. de Roo , M.B. Green, G. Papadopoulos, 
  P.K. Townsend,
 {\it Duality of Type II 7 Branes and 8 Branes},
  Nucl.Phys. {\bf B470}:113-135,1996; 
 hep-th/9601150.}     
  \REF\cfgp{E. Cremmer, S. Ferrara, L. Girardello, A. Van
Proeyen, {\it Yang-Mills Theories with Local Supersymmetry: Lagrangian,
Transformation Laws and Superhiggs Effect},  Nucl.Phys. {\bf B212}:413,1983.}
\REF\bw{
 J. Bagger, {\it Supersymmetric Sigma Models} 
Lectures given at Bonn-NATO Advanced Study Inst. on Supersymmetry, Bonn,  
Germany.
E. Witten, J. Bagger, {\it
Quantization of Newton's Constant in Certain Supergravity Theories},  
 Phys.Lett.{\bf B115}:202,1982. }
 \REF\lambert{ P.S. Howe, N.D. Lambert , P.C. West,
 {\it A New Massive Type IIA Supergravity from Compactification},
 Phys. Lett. {\bf B416}:303, 1998; 
 hep-th/9707139.}

\Pubnum{ \vbox{ \hbox{}\hbox{} } }
\pubtype{}
\date{July, 2001}
\titlepage
\title{ Magnetic Cosmic Strings of $\cN=1, D=4$ Supergravity 
with  Cosmological Constant}
\author{J. Gutowski}
\address{Department of Physics,\break Queen Mary College, 
\break Mile End,\break London E1 4NS}
\andauthor{G. Papadopoulos }
\address{Department of Mathematics,\break King's College London, \break
Strand,\break
London WC2R 2LS}

\abstract {We find a new class of cosmic string solutions with non-vanishing
 magnetic flux
 of $\cN=1$, $D=4$ supergravity with a
cosmological constant and coupled to any number of Maxwell and scalar
multiplets. We show that  these magnetic cosmic string solutions preserve 
$1/2$ of  supersymmetry. We give an explicit example of such a solution
for which the complex scalars are constant and the spacetime is smooth with 
  topology $R^{1,1}\times S^2$. Two more examples are explored
for which a complex scalar field takes values in $\bC P^1$ and  in
$SL(2,\bR)/U(1)$. }

\vskip 1.0 cm
\endpage
\pagenumber=2
\font\mybb=msbm10 at 12pt
\def\bb#1{\hbox{\mybb#1}}

\def\C{\mkern1mu\raise2.2pt\hbox{$\scriptscriptstyle|$}\mkern-7mu{\rm C}}

\def\log{{\rm log}}

\def\a{\alpha}
\def\pd{\partial_}

\def\b{\beta}
\def\g{\gamma}
\def\s{\sigma}

\def\e{\epsilon}

\def\cN {{\cal {N}}}

\sequentialequations

The investigation of the properties of $D=10$ and $D=11$
supergravities has given new insights in the non-perturbative nature
of strings. This has been mostly achieved by studying the
 supersymmetric brane solutions of $D=10$ and $D=11$
 supergravities as well as their reductions to four and five
dimensions.
Conversely  certain solutions of D=4 or D=5
supergravity theories when lifted to ten and eleven dimensions
were given an interpretation in terms of brane configurations. 
 One such class of solutions are the cosmic
strings of $D=4$ gravity coupled to scalars (see for example [\cg] and [\vy]). 
These solutions can be embedded within
various supergravity theories without a scalar potential
or a cosmological constant and typically preserve
$1/2$ of the supersymmetry. In particular the solution given in
[\vy] when lifted to type IIB supergravity becomes the D7-brane [\gg].
Reducing these solutions to three dimensions along the
spacelike worldvolume direction of the string,  they become
particle solutions of three dimensional gravity and 
 exhibit conical singularities at the position
of the particles. The total mass of the solution is identified
with the deficit angle of the spacetime at infinity. If appropriate
number of cosmic strings are included in the spacetime, then
the spacetime  closes.

In some of the above investigations, the $D=4$ and $D=5$ supergravities
considered do not have a scalar potential. A scalar potential can
nevertheless appear because of non-perturbative effects.
Their presence is desirable because supersymmetry is broken
at low energies. This is for example achieved by giving
an expectation value to some fields. In the presence of Maxwell
fields one can use 
the Fayet-Iliopoulos mechanism. In a gravitational setting, this leads
to the addition of a  cosmological constant $\Lambda$ in the gravity action
which is the length of the $D$-term. In such a case, the \lq\lq vacuum" of
theory is not supersymmetric. However it is possible to find supersymmetric
solutions for which the effect of the cosmological
constant in the Killing spinor equations is negated by other fields
 typically scalars or Maxwell fields.
This situation is reminiscent of that of IIA massive supergravity [\rom] where
although the \lq\lq vacuum" of the theory is not supersymmetric, there is
a supersymmetric solution which is the D8-brane [\wipo, \ptb]. In this case 
the cosmological constant is balanced against a contribution from the dilaton.

In this paper, we shall investigate   ${\cal N}=1, D=4$ supergravity theory
with a cosmological constant coupled to any number 
of  Maxwell and  scalar multiplets [\cfgp, \bw].
It is well known that the sigma model manifold in which
the complex  scalar fields take values is Hodge, ie it is K\"ahler
and the K\"ahler form $\Omega$ represents an integral
cohomology class\foot{It has been argued that the latter property  leads to
the quantization of the Newton constant [\bw].}.
We shall find that if the Maxwell fields vanish, then the theory
breaks all supersymmetry. However in the presence of Maxwell fields,
we shall show that there is a solution which preserves $1/2$
of the supersymmetry. The solution exhibits $1+1$ Poincar\'e invariance
and has non-vanishing magnetic flux, ie it is a magnetic cosmic string. 
One of the novel features of this solutions is that as
a consequence of the Killing spinor equations, the Maxwell fields
obey the Hermitian-Einstein equations with respect to the K\"ahler
form of the spacetime metric; 
 D-term appears naturally in this equation. We shall find that there are 
 smooth solutions. One example involves all the scalars to be constant.
In this example for an appropriate choice of parameters, the
spacetime metric is smooth and topologically $\bR^{1,1}\times S^2$.
We also give the flux of the Maxwell field through the two-sphere. 
Two more examples will be explored,
in one  a complex scalar takes values in $\bC P^1$ and in the
other it takes values in $SL(2, \bR)/U(1)$.

For our purpose, we shall consider the special case
of ${\cal N}=1$ $D=4$ supergravity action with $n$ 
vector (abelian) $A^a$
and $m$ chiral multiplets $z^i$ up to fermion terms given by
$$
\eqalign{
L&= \sqrt{-g} \big[{1\over2} R(g)-{1\over4} {\rm Re}h_{ab} F^a_{MN} F^b{}^{MN} 
+{1\over 4} \iim h_{ab}{}^\star F^{a MN}F{}^b{}_{MN} 
\cr &
-\g_{i\bar j}\partial_M z^i
\partial^M z^{\bar
j}-{1 \over 2}{\rm Re}h^{ab}D_a D_b \big]\ ,}
\eqn\ggone
$$
where $g$ is the spacetime metric\foot{Note that $[\nabla_M, \nabla_N] V^L=
R_{MN}{}^L{}_K V^K$ and $R_{MN}=R_{LM}{}^L{}_N$.}, 
$\g$ is the metric on the chiral multiplets
sigma model manifold which is K\"ahler, $h$ is a holomorphic function 
of the chiral multiplets  ($({\rm Re}h_{ab})^{-1}={\rm Re}h^{ab}$) and
$$
F^a_{MN}=\partial_M A_N^a-\partial_N A_M^a\ ;
\eqn\ggtwo
$$
$M,N=0,1,2,3$ are spacetime indices, $i,j=1,\dots,m$ are sigma model
manifold indices and $a,b=1,\dots,n$ are gauge indices. This Lagrangian
has a cosmological constant given by the length of the D-term;
$D_a$ is taken to be {\it constant}.

The supersymmetry transformations can be written in terms
of a real four-component Majorana spinor $\e$ as
$$
\eqalign{
2 \big( \pd{M} +{1 \over 4} \omega_{M {\underline A} {\underline B}}
\Gamma^{{\underline A} {\underline B}} \big) \e
- \Gamma^5 {\rm Im} (K_i \partial_M z^i) \e - \Gamma^5 A{}^a{}_M 
 D_a \e=0\ ,}
 \eqn\ggthree
$$
$$
\big( -{1 \over 2} F{}^a{}_{MN} \Gamma^{MN} + \Gamma^5 D^a \big) \e=0
\eqn\ggfour
$$
and
$$
\eqalign{
\big( {\rm Re} (\partial_M z^i) - \Gamma^5 
{\rm Im} (\partial_M z^i) \big) \Gamma^M \e =0\ ,}
\eqn\ggfive
$$
where underlined indices ${\underline A}, \ {\underline B}$
 denote tangent frame
indices and $\Gamma^5 = \Gamma^{\underline{0}}
 \Gamma^{\underline{1}} \Gamma^{\underline{2}}
\Gamma^{\underline{3}}$ and gauge indices are raised with ${\rm Re} h^{ab}$.

The field equations of the above Lagrangian are as follows:
\item{(1)} The Einstein equations are
$$
  G_{MN} + {1\over2} g_{MN} D_a D^a=T_{MN}\ ,
  \eqn\fone
  $$
  where
  $$
  \eqalign{
T_{MN}=& {\rm Re} h_{ab} F{}^a{}_{ML} F{}^b{}_N{}^L +2
 \g_{i {\bar j}} \partial_{(M} z^i \partial_{N)} z^{\bar j} 
\cr &-  g_{MN} \big( {1 \over 4} 
{\rm Re} h_{ab} F{}^a{}_{LD} F{}^{b LD} +
\g_{i {\bar j}}\, \partial_{L} z^i \partial^{L} z^{\bar j} 
 \big)\ .}
 \eqn\enmom
$$

\item{(2)} The gauge equations are

$$
\eqalign{
\pd{M} \big( \sqrt{-g} \big[ {\rm Re} h_{ab} F^{b MN}
- \iim h_{ab} {}^\star F^{b MN} \big] \big) \  =0}
\eqn\ftwo
$$

\item{(3)} The scalar equations by varying $z^\ell$ are
$$
\eqalign{
 -{1 \over 8} \pd{\ell} &h_{ab} F{}^a{}_{MN} F{}^{b MN} - 
 {1 \over 2} \pd{\ell} (D^a D_a) -
 {i \over 8} \pd{\ell} h_{ab} {}^{\star} F^{a MN} F{}^b{}_{MN}
 \cr &
+ \g_{\ell \bar j} {\tilde \nabla}_M \partial^M z^{\bar j}  =0\ ,}
\eqn\fthree
$$ 
where 
$$
{\tilde{\nabla}}_M \partial^N z^{\bar i} = \nabla_M \partial^N z^{\bar i} 
+ \Gamma^{\bar i}{}_{\bar j \bar k} \pd{M} z^{\bar j} \partial^N z^{\bar k}\ ,
\eqn\ggsix
$$
and $D^a={\rm Re}h^{ab} D_b$.
Taking the conjugate of this equation, one obtains the field equation for 
$z^{\bar \ell}$.

To simplify further the equations we set
$h_{ab}=\pm \delta_{ab}$. If one takes $F^a=0$, $z^i={\rm const}$, then
the field equations reduce to 
$$
G_{MN}\pm {1\over2} |D|^2 g_{MN}=0\ ,
\eqn\ggseven
$$
ie the spacetime is an Einstein manifold, where $|D|^2=|D_a D^a|$. 
The maximally symmetric
solutions with $|D|^2\not=0$  are either de Sitter ($D_a D^a>0$) or 
anti-de Sitter ($D_a D^a<0$) space  depending on the sign
of the cosmological constant $\Lambda={1\over2} D_a D^a$. 
In both cases the solutions are not supersymmetric.
This is most easily seen by looking at the Killing spinor 
equations associated with the gaugino. 
To construct supersymmetric solutions,
one has to introduce a non-trivial Maxwell field to balance
the D-term in the gaugino Killing spinor equations.
It is also worth  mentioning that if the inner product
 ${\rm Re}h$ is {\sl positive definite}, then
the energy of the the Maxwell fields {\sl  does}  obey the
weak energy condition while if it is {\sl negative definite}
it {\sl does not}.

Suppose that the magnetic cosmic string solution that 
we are seeking lies in directions
$0,1$ with transverse directions $2,3$. We write the
 ansatz
$$
\eqalign{
ds^2 = -dt^2 +d\s^2 +B^2(x,y) (dx^2+dy^2)
\cr
z^i = z^i(x,y)
\cr
A^a = A^a{}_x(x,y) dx + A^a{}_y(x,y) dy\ .}
\eqn\pone
$$
Substituting this into the Killing spinor equations,
we find
$$
\eqalign{
2 \pd{x} \e + \pd{y} \log B  \Gamma_\ubx \Gamma_\uby \e
- \Gamma^5 \iim \big( K_i \pd{x} z^i +i D_a A^a{}_x \big) \e=0
\cr
2 \pd{y} \e  - \pd{x} \log B \Gamma_\ubx \Gamma_\uby \e
- \Gamma^5 \iim \big( K_i \pd{y} z^i +i D_a A^a{}_y \big) \e=0\ ,}
\eqn\ptwo
$$
$$
\big( -B^{-2} F^a{}_{xy} \Gamma_\ubx \Gamma_\uby + 
\Gamma^5 D^a \big) \e =0
\eqn\pthree
$$
and
$$
\big( \rre \partial_{x} z^i - \Gamma^5 \iim \partial_{x} z^i \big)
 \Gamma_\ubx \e
+ \big( \rre \partial_{y} z^i - \Gamma^5 \iim \partial_{y} z^i \big)
 \Gamma_\uby \e =0\ ..
 \eqn\pfive
$$
To solve the above Killing spinor equations, we impose the
condition\foot{In fact we can also impose the condition 
$\Gamma^5 \Gamma_\ubx \Gamma_\uby \e =  \e$ but this will lead to
anti-holomorphic solutions.}
$$
\Gamma^5 \Gamma_\ubx \Gamma_\uby \e = - \e\ .
\eqn\susypro
$$
Observe that this is equivalent to 
$\Gamma^{\underline 0} \Gamma^{\underline 1}\e=\e$.
Next introduce complex coordinates $u=x+iy$. Using the
condition \susypro, the Killing spinor equations become
$$
\eqalign{
2\partial_u\e+\big[i\partial_u (\log B+{1\over2} K)
-D_a A^a_u\big]\Gamma^5\e&=0
\cr
-2iF^a_{u\bar u}&=B^2 D^a
\cr
\partial_{\bar u} z^i&=0\ .}
\eqn\kse
$$
The last equation implies that $z^i$ is holomorphic.
The second equation in \kse\ can be rewritten as
$$
F^a=D^a \Omega\ ,
\eqn\ke
$$
where $\Omega$ is the K\"ahler form associated with the curved
part of the spacetime metric. So it is recognized 
as the Hermitian-Einstein
equation for the connection $A$.
To proceed, since the curvature $F$ of $A$ is a (1,1)-form, we can always
locally write
$$
A^a_u=i \partial_u Y^a\ ,
\eqn\con
$$
where we have used the gauge transformations to set $Y^a$ real.
In order for the first equation in \kse\ to have solutions, the connection
associated with the parallel transport should be trivial. For this
we have
$$
\log B+{1\over2}K-D_a Y^a=f(u)+\bar f(u)\ ,
\eqn\fc
$$
where $f$ is an arbitrary locally defined holomorphic function.
The Killing spinors then are given by
$$
\epsilon=e^{{\rm Im}f \Gamma^5}\epsilon_0\ ,
\eqn\psix
$$
where $\epsilon_0$ is a constant spinor satisfying the projection 
$\Gamma^{\underline 0} \Gamma^{\underline 1}\e_0=\e_0$. Therefore
the solution preserves $1/2$ of supersymmetry.

To solve the rest of the Killing spinor equations in \kse, we have
to determine $B$ and $Y^a$ from \ke\ and \fc. Observe
that without loss of generality, we can set
$$
Y^a=Y D^a\ .
\eqn\cona
$$
To see this decompose $Y^a$ into parallel and perpendicular  parts with 
respect to $D^a$. If the only non-vanishing components of $Y^a$ are the
 perpendicular ones, then it is easy to see that $F^a=0$ and consequently
 from \ke\ $D^a=0$. Thus the solution reduces to that of cosmic strings of 
 [\cg, \vy].
 
 Next if the only non-vanishing component of $Y^a$ is parallel to $D^a$,
 we set $Y^a=Y D^a$.
 Then \ke\ and \fc\ equations imply
 $$
 \eqalign{
 B&=e^{-{1\over2}K+2 {\rm Re}f+ 2\Lambda Y}
 \cr
 \partial^2 Y&= -e^{4\Lambda Y} e^{-K+4 {\rm Re}f}\ ,}
 \eqn\fs
 $$
 where $\partial^2=\partial_x^2+\partial_y^2$.
 It also turns out that the above two equations  imply
  the field equations \fone, \ftwo\ and \fthree.
 
 In the mixed case where $Y^a$ has both parallel and perpendicular
 components with respect to $D^a$,  \ke\ and \fc\ imply 
 that the perpendicular components
 should vanish up to the real part of a locally 
 defined holomorphic function. In particular the perpendicular
 components of $Y^a$ do not contribute either in the field strength
 $F^a$ or in the spacetime metric. Therefore the mixed case reduces
 to that for which $Y^a$ is parallel to $D^a$.

 The simplest possible case to investigate 
 is the one with $z^i={\rm const}$
 and $f=0$.  Under these assumptions the
  second equation \fs\ becomes
 the Liouville equation. After a constant 
 shift in $Y$ to absorb $K$,
 the equation can be rewritten as
 $$
 \partial^2 Y= -e^{4\Lambda Y}\ .
 \eqn\ll
 $$ 
 There are two cases to consider depending on whether  $\Lambda$
 positive or negative. 
 We shall first consider the case where $\Lambda>0$.
  A spherically symmetric solution is
 $$
 Y(r)={1\over 4\Lambda} \log\big[{\alpha^2 \over 8\Lambda r^2}
  \cosh^{-2}\big({\alpha\over2} (\log r-\beta)\big)\big]\ .
  \eqn\qone
  $$
  Then the spacetime metric and the Maxwell field strengths  are given by
  $$
  \eqalign{
  ds^2&=-dt^2+d\s^2+ {\alpha^2 \over 8\Lambda r^2}
  \cosh^{-2}\big({\alpha\over2} (\log r-\beta)\big) (dr^2+r^2 d\theta^2)
  \cr
  F^a&= {\alpha^2\over 8\Lambda r} \cosh^{-2}\big({\alpha\over2}
  (\log r-\beta)\big) D^a dr\wedge d\theta\ ,}
  \eqn\qtwo
  $$
where $\alpha$ and $\beta$ are integration constants.
The asymptotic behaviour of 
the metric as $r\rightarrow \infty$ is
$$
 ds^2\sim -dt^2+d\s^2+ {e^{\a \b} \alpha^2 \over 2\Lambda} r^{-(2+\alpha)}
  (dr^2+r^2 d\theta^2)\ .
  \eqn\qsix
 $$
 This asymptotic  expression for the metric gives
 a deficit angle $\delta=\pi (2+\alpha)$.
 On the other hand the behaviour of the metric as $r\rightarrow 0$ is
$$
 ds^2\sim -dt^2+d\s^2+ {\alpha^2 e^{-\a \b} \over 2\Lambda} r^{-2+\alpha}
  (dr^2+r^2 d\theta^2)
  \eqn\qsev
  $$   
Changing coordinates as $v={1\over r}$, the metric becomes
$$
 ds^2\sim -dt^2+d\s^2+ {\alpha^2 e^{-\a \b} \over 2\Lambda} v^{\alpha-6}
  (dv^2+v^2 d\theta^2)\ .
  \eqn\qeigth
  $$
This gives another asymptotic region with deficit angle
$\delta=\pi (6-\alpha)$. For most values of $\alpha$, the
spacetime has conical singularities. In particular for $\alpha=0$ and so
$\delta=2\pi$ which is the deficit angle for cylindrical asymptotic
behaviour,
the metric degenerates.  However for $\alpha=2$      
 the spacetime is {\sl smooth} and topologically $\bR^{1,1}\times S^2$.
The fluxes, $\Phi^a$, of the Maxwell fields through the two sphere are
$$
\Phi^a=
\int_0^{2\pi}d\theta \int_0^\infty dr\, F^a_{r\theta}=
{2\pi \over \Lambda} D^a\ .
\eqn\don
$$

Alternatively, we can consider the case that $\Lambda<0$.
The solution of the equation \ll\ in this case is
$$
Y= {1\over 4\Lambda} \log\big[ {\a^2\over 8 |\Lambda| r^2} 
\cos^{-2} \big({\a\over2}(\log r-\b)\big)\big]\ .
\eqn\dtwon
$$
This leads to the metric
$$
ds^2=-dt^2+d\s^2+{\a^2 \over 8 
|\Lambda| r^2}\cos^{-2} \big({\a\over2}(\log r-\b)\big)
(dr^2+r^2 d \theta^2)\ .
\eqn\ddthree
$$
After changing coordinates $r=e^{\b + v}$ , we find
$$
ds^2=-dt^2+d\s^2+{\a^2 \over 8 |\Lambda|} \cos^{-2} 
({\a v \over 2}) (dv^2+d \theta^2)\ .
\eqn\dsone
$$
This metric is regular at the points that $\cos ({\a v \over 2})$
 vanishes. This can easily
been seen by introducing co-ordinates $v= {(2m+1) \pi \over \a} + \rho$ 
for $m \in \bZ$. 
Then in the neighbourhood of $v={(2m+1) \pi \over \a}$ the metric becomes
$$
ds^2 \sim -dt^2+d\s^2 +{1 \over 2 |\Lambda| \rho^2}(d \rho^2+d \theta^2)\ ,
\eqn\ssone
$$
ie the geometry is $\bR^{1,1}\times H^2$, where $H^2$ is the two-dimensional
hyperbolic space; $\theta$ is now not  periodic.
However the metric is singular as $v\rightarrow \pm \infty$.

Next consider the example for which there is an active scalar $z$ which
takes values in $\bC P^1$. The  K\"ahler potential is $K=2q \log (1+z \bar 
z)$ and we choose  $q \in \bZ^+$.
The solution for $z$ is
$$
z(u) = \sum_{r=1}^N {u-a_r \over u-b_r}\ .
\eqn\sstwo
$$
This solution is the same as that for the complex scalar in the
case of cosmic strings without a cosmological constant given in
[\cg]. Similarly, we take
$$
e^{4 \rre f}  = {1 \over \prod_{i=1}^N \mid u - b_r \mid^{4q}}\ .
\eqn\ssthree
$$
Obtaining an analytic solution of \fs\ in terms of $Y$ is somewhat awkward. 
It is however possible to
examine the asymptotic behaviour of the solutions. 
In particular, as
 $\mid u \mid \rightarrow \infty$, one finds that
 $e^{-K+4 \rre f} \rightarrow {1 \over 2^q}r^{-4qN}$.
 So, asymptotically \fs\ simplifies to
$$ 
e^{-2 D_a D^a Y} \partial^2 Y = -{1 
\over 2^q} r^{-4qN}\ .
\eqn\ssfour
 $$ 
This can be  solved by taking 
$$ 
Y = {1 \over 4\Lambda} \log \big({2^{q-3}\a^2 \over \Lambda
r^{2-4qN}} \cosh^{-2} ({\a \over 2} ( \log r - \b)) \big) 
\eqn\ssfive
$$ 
where $\a$ and $\b$ are integration constants and $\Lambda>0$. 
As $r \rightarrow \infty$  the metric becomes
$$
 ds^2\sim -dt^2+d\s^2+ {2^{q-1} e^{\a \b} \alpha^2 \over \Lambda} 
 r^{-(2+\alpha)}(dr^2+r^2 d\theta^2)\ .
  \eqn\sssix
 $$
This asymptotic metric is identical to that we have found
in the previous  example where all the complex scalars were taken to be
constant. Therefore the deficit angle at infinity is
$\delta=\pi (2+\alpha)$ and so the spacetime closes smoothly at this
asymptotic region to a two-sphere for $\alpha=2$.   

In the third example, we take the complex scalar $z$ to take
values $SL(2, \bR) /U(1)$. In fact we take $z$ to take values
in the fundamental domain of the $SL(2,\bZ)$ acting on
$SL(2, \bR) /U(1)$ with modular transformations.
This is similar to the case of stringy cosmic strings considered in
[\vy].
 The K\"ahler potential of $SL(2, \bR) /U(1)$
  is $K = - \log \iim z$. The complex scalar is determined implicitly
  via the equation
$$
j(z) = {P(u) \over Q(u)}\ ,
\eqn\zone
$$
where $P$, $Q$ are polynomials, and $j(z)$ denotes the
 modular $j$-function. In this case we take
$$
e^{-K +4 \rre f} = \iim z |\eta(z)|^4 
\mid \prod_{r=1}^N (u - u_r)^{-{1 \over 12}} \mid^2\ ,
\eqn\ztwo
$$
where $\eta (z) = e^{\pi i z \over 12} \Pi_{r>0} (1- e^{2 \pi i r z})$ denotes
the Dedekind $\eta$-function and $N={\rm max}({\rm deg}P, {\rm deg}Q)$.
Again, we have not been able to find an explicit solution for $Y$ in \fs.
However, we can determine the asymptotic behaviour of the solution
at infinity. As $r \rightarrow \infty$,  we find that $e^{-K+4 \rre f} 
\rightarrow r^{-{N \over 6}}$. In this case,  $Y$ is given by
$$
 Y = {1 \over 4\Lambda} \log \big({\a^2 \over 8\Lambda
r^{2-{N\over6}}} \cosh^{-2} ({\a \over 2} ( \log r - \b)) \big)\ , 
\eqn\zthree
$$ 
where $\a$ and $\b$ are integration constants and $\Lambda>0$. 
For this solution of $Y$, the metric asymptotically is 
$$
 ds^2\sim -dt^2+d\s^2+ {e^{\a \b} \alpha^2 \over 2\Lambda} r^{-(2+\alpha)}
  (dr^2+r^2 d\theta^2)\ .
  \eqn\zfour
 $$
So the deficit angle at this asymptotic
region  is  $\delta=\pi(2+\a)$  and  the spacetime closes smoothly to
a two-sphere for $\alpha=2$.

It is natural to ask the question whether the magnetic cosmic string
solutions we have found here can be lifted to ten or eleven dimensions.
For this to happen, the reduction of the associated D=10 or D=11
supergravity to four dimensions should give a cosmological term in
four-dimensions. Standard compactifications do not have this property
but this may  change if certain form field strengths in D=10 or D=11
supergravity receive non-vanishing expectation values. Alternatively
one may be able to lift these solutions to 
D=10 supergravity theories with cosmological
constant like that of massive IIA or the one 
found in [\lambert].

\vskip 1cm
\noindent{\bf Acknowledgments:}  We would like to thank Jose M Figueroa-O'Farrill
for helpful discussions. J.G. is supported by a EPSRC postdoctoral grant.
G.P. is supported by a University Research
Fellowship from the Royal Society. This work is partially supported by SPG grant
PPA/G/S/1998/00613.
\refout
\end